\documentclass[onecolumn]{mn2e}
\newif\ifAMStwofonts

\def\pmb#1{\mbox{\boldmath$#1$}}
\def\gtsim {>\kern-1.2em\lower1.1ex\hbox{$\sim$}}
\def\ltsim {<\kern-1.2em\lower1.1ex\hbox{$\sim$}}
\def\gtsim {>\kern-1.2em\lower1.1ex\hbox{$\sim$}}
\def\ltsim {<\kern-1.2em\lower1.1ex\hbox{$\sim$}}
\def\ref{\hangindent=1pc \hangafter=1 \noindent}

\def\be{\begin{equation}}
\def\ee{\end{equation}}

\usepackage{epsfig}
\begin{document}



\title{Axisymmetric toroidal modes of magnetized neutron stars}
\author[U. Lee]{Umin Lee$^1$\thanks{E-mail: lee@astr.tohoku.ac.jp}
\\$^1$Astronomical Institute, Tohoku University, Sendai, Miyagi 980-8578, Japan}

\date{Typeset \today ; Received / Accepted}
\maketitle


\begin{abstract} 
We calculate axisymmetric toroidal modes of magnetized neutron stars
with a solid crust.
We assume the interior of the star is threaded by a poloidal magnetic field 
that is continuous at the surface with the outside dipole field whose strength $B_p$
at the magnetic pole is $B_p\sim10^{16}$G.
Since separation of variables is not possible for oscillations of 
magnetized stars, we employ finite series expansions of the perturbations
using spherical harmonic functions to represent 
the angular dependence of the oscillation modes.
For $B_p\sim 10^{16}$G, we find distinct mode sequences, in each of which the
oscillation frequency of the toroidal mode
slowly increases as the number of radial nodes of the eigenfunction increases.
The frequency spectrum of the toroidal modes for $B_p\sim10^{16}$G is largely different from
that of the crustal toroidal modes of the non-magnetized model, although the frequency ranges
are overlapped each other.
This suggests that 
an interpretation of the observed QPOs based on the magnetic toroidal modes
may be possible if the field strength of the star
is as strong as $B_p\sim10^{16}$G.
\end{abstract}


\section{Introduction}

Recent discovery of quasi-periodic oscillations (QPOs) of magnetar candidates
is one of the observational manifestations of global oscillations of neutron stars.
Israel et al (2005) detected QPOs 
of frequencies $\sim18$, $\sim 30$ and $\sim$92.5Hz 
in the tail of the SGR 1806-20 hyperflare observed December 2004, and suggested 
that the 30Hz and 92.5Hz QPOs
could be caused by seismic vibrations of the neutron star crust (see, e.g., Duncan 1998).
Later on, in the hyperflare of SGR 1900+14 detected August 1998, 
Strohmayer \& Watts (2005) found QPOs 
of frequencies 28, 53.5, 84, and 155 Hz, and claimed that the QPOs
could be identified with the low $l$ fundamental toroidal torsional modes 
of the solid crust of the neutron star.
These recent discoveries of QPOs in the giant flares 
of Soft Gamma-Ray Repeaters
SGR 1806-20 (Israel et al 2005, Watts \& Strohmayer 2006, Strohmayer \& Watts 2006) and 
SGR 1900+14 (Strohmayer \& Watts 2005)
have made promising asteroseismology for magnetars,
neutron stars with an extremely strong magnetic field
(see, e.g., Woods \& Thompson 2006 for a review of SGRs).

It is currently common to identify these QPOs with seismic vibrations caused by
crustal toroidal modes of the neutron stars, since the frequency range of 
the modes overlap that of the observed
QPOs and from the energetics point of view the crustal toroidal modes would be most easily
excited to observable amplitudes by spending a least amount of available energies, 
for example, those released in magnetic field restructuring (e.g., Duncan 1998). 
Although the interpretation based on crustal torsional modes looks promising,
we need detailed theoretical analyses of oscillations of magnetized neutron stars
so that we could get information of physical conditions of the stars through
the confrontation between theoretical modelings and observations.
This is particularly true for high frequency QPOs 
(e.g., 625Hz QPO, Watts \& Strohmayer 2006; 1835Hz QPO and less significant QPOs at
720 and 2384 Hz in SGR 1806-20, Strohmayer \& Watts 2006), 
since there exist classes of modes other than the crustal toroidal modes that can 
generate the frequencies observed.

The presence of a magnetic field makes it possible for toroidal modes to exist in 
a fluid star even without rotation as does the presence of the shear modulus in the solid crust.
In this paper we are interested in axisymmetric toroidal modes since
axisymmetirc toroidal and spheroidal modes are decoupled for a poloidal field
when the star is non-rotating.
For non-axisymmetric modes, the toroidal and spheroidal components are
coupled even without rotation and hence the modal analyses of magnetized stars 
would be much more complicated.

Theoretical calculations of toroidal modes of strongly magnetized neutron stars
have been carried out by several authors, including
Piro (2005), Glampedakis et al (2006), Sotani et al (2006, 2007), and Lee (2007).
The analyses by Piro (2005), Glampedakis et al (2006), and Lee (2007) assume Newtonian gravity, 
while
those by Sotani et al (2006, 2007) use general relativistic formulation.
Although the studies by Piro (2006) and Lee (2007) ignore the effects of magnetic fields 
in the fluid core, those by Glampedakis et al (2006) and Sotani et al (2006, 2007) consider 
magnetic waves propagating in the fluid core, assuming the core is threaded by a magnetic field of
substantial strength.
Besides the differences mentioned above, most of the authors except Lee (2007) represent the
angular dependence of the oscillations by a single spherical
harmonic function $Y_l^m(\theta,\phi)$.
Since the shear modulus in the crust dominates the magnetic pressure
in most parts of the crustal regions for a dipole field of strength $B_p\ltsim 10^{15}$G,
this treatment may be justified so long as the crustal toroidal modes are well decoupled 
from the fluid core.
But, if the torsional waves in the crust are strongly coupled with magnetic waves in the core,
the treatment may not be justified
because the angular dependence of the magnetic waves
in the fluid core cannot be correctly represented by a single spherical harmonics.
For example, the analysis by Reese, Fincon, \& Rieutord (2004) of toroidal modes in a fluid shell
have employed finite series expansions of long length for the perturbations.

In this paper, using the method of series expansions of perturbations
we calculate toroidal modes of a strongly magnetized neutron star having a fluid core
and a solid crust, where the entire interior is assumed to be
threaded by a poloidal magnetic field.
We employ two different sets of oscillation equations, one for fluid regions and 
the other for the solid crust, and solutions in the solid and fluid regions 
are matched at the interfaces between them
to obtain an entire solution of a mode.
The method of calculation we employ is presented in \S 2, and
the numerical results are given in \S 3,
and conclusions are in \S 4.

\section{Method of Solution}

\subsection{Magnetic Fields in the Interior}

We employ an inner magnetic field of Ferraro (1954) type.
For a magnetized star in hydrostatic equilibrium, Ferraro (1954) assumed
an interior poloidal field given by 
\be
B_r={1\over r^2\sin\theta}{\partial\over\partial\theta}U, \quad
B_\theta=-{1\over r\sin\theta}{\partial\over\partial r}U,
\ee
where $U$ is a scalar function, and looked 
for a particular solution to the partial differential equation:
\be
{1\over r\sin\theta}\left[{\partial^2U\over\partial r^2}
+{\sin\theta\over r^2}{\partial\over\partial\theta}
\left({1\over\sin\theta}{\partial\over\partial\theta}U\right)\right]=\kappa r^2\sin^2\theta,
\ee
where $\kappa$ is a constant to be determined.
Assuming the scalar function $U$ is given by
\be
U=\left(C_1r^2+C_2r^4\right)\sin^2\theta,
\ee
and imposing the condition that the interior field is continuous at the stellar surface 
with the outside dipole field, Ferraro (1954) obtained
\be 
C_1=-{5\over 4}B_p, 
\quad C_2={\kappa\over 10}={3\over 4}{B_p\over R^2},
\ee
where $B_p$ denotes the strength of the magnetic dipole field at the pole, and $R$ is the radius of the star.
For modal analysis shown below, we use this equilibrium magnetic field configuration with $B_p$ being
a parameter.

\subsection{Oscillation Equations}

We assume the temporal and azimuthal angular dependence of perturbations is given by
a single factor $e^{i\left(m\phi+\omega t\right)}$ for oscillations of 
magnetized, non-rotating stars, where
$\omega$ is the oscillation frequency in the inertial frame, 
and $m$ denotes the azimuthal wave number.
The linearized basic equations governing axisymmetric ($m=0$)
toroidal modes propagating in the solid crust of the magnetized star may be given by
\be
-\omega^2\xi_\phi={1\over\rho}\left[\nabla\cdot\pmb{\sigma}^\prime\right]_\phi
+{1\over 4\pi\rho}\left[\left(\nabla\times\pmb{B}^\prime\right)\times\pmb{B}\right]_\phi,
\ee
\be
B^\prime_\phi=\left[\nabla\times\left(\pmb{\xi}\times\pmb{B}\right)\right]_\phi.
\ee
In equation (5), $\pmb{\sigma}^\prime$ denotes the Euler perturbation of the stress tensor and is
obtained from the Lagrangian perturbation defined in Cartesian coordinates by
\be
\delta\sigma_{ij}=(\Gamma_1pu)\delta_{ij}+2\mu (u_{ij}-{1\over 3}u\delta_{ij})
\ee
with $u_{ij}$ being the strain tensor defined by
\be
u_{ij}={1\over 2}\left({\partial\xi_i\over\partial x_j}+{\partial\xi_j\over\partial x_i}\right),
\ee
where $\delta_{ij}$ denotes the Kronecker delta, $\mu$ is the shear modulus, and 
$u=\sum_{l=1}^3u_{ll}$ ( see, e.g., McDermott et al 1988,
and Lee \& Strohmayer 1996).
The linearized equations for toroidal modes in the fluid regions are obtained by simply dropping
the term $\rho^{-1}\nabla\cdot\pmb{\sigma}^\prime$ in equation (5).

Since separation of variables is not possible for oscillations
of magnetized stars, we employ series expansions of a finite length $j_{\max}$ for the displacement vector
$\pmb{\xi}$ and the perturbed magnetic field $\pmb{B}^\prime$
using spherical harmonic functions $Y_l^m(\theta,\phi)$ for a given $m$.
We have for axisymmetric toroidal modes
\be
{\xi_\phi\over r}=-\sum_{j=1}^{j_{\rm max}}T_{l^\prime_j}(r){\partial\over\partial\theta}Y_{l^\prime_j}^m(\theta,\phi)e^{i\omega t},
\ee
\be
{B^\prime_\phi\over B_0}=-\sum_{j=1}^{j_{\rm max}}b^T_{l_j}(r){\partial\over\partial\theta}Y_{l_j}^m(\theta,\phi)e^{i\omega t},
\ee
where $l_j=|m|+2(j-1)$ and $l^\prime_j=l_j+1$ for even modes, and 
$l_j=|m|+2j-1$ and $l^\prime_j=l_j-1$ for odd modes, respectively, and $j=1,~2,~3,~\cdots, ~j_{\rm max}$,
and $B_0$ is a normalizing constant, which is set equal to $B_p$.
Note that, for axisymmetric toroidal modes with $m=0$, we have to redefine $l_j$ and $l^\prime_j$ as
$l_j=2j$ and $l^\prime_j=l_j-1$ for even modes, $l_j=2j-1$ and $l^\prime_j=l_j+1$ for odd modes for
$j=1,~2,~3,~\cdots,~j_{\rm max}$.
Most of the numerical results shown below are obtained for $j_{\rm max}=20$.
In this convention, the angular pattern of $B^\prime_\phi$ (of $\xi_\phi$) at the stellar surface is
symmetric (anti-symmetric) about the equator for even modes and it is anti-symmetric (symmetric) for odd modes.

If we use as dependent variables the vectors $\pmb{t}$ and $\pmb{b}^T$ defined by
\be
\pmb{t}=\left(T_{l^\prime_j}(r)\right), \quad \pmb{b}^T=\left(b^T_{l_j}(r)\right),
\ee
the oscillation equations for fluid regions are given by
\be
{B_1\over B_0}\pmb{Q}_1\hat{\pmb{C}}_0r{d\over dr}i\pmb{t}=\hat{\pmb{C}}_1i\pmb{b}^T
-\left[{B_1\over B_0}\left(2+{d\ln B_r\over d\ln r}\right)\pmb{Q}_1
-{B_2\over B_0}\pmb{C}_1\right]\hat{\pmb{C}}_0i\pmb{t},
\ee
\be
{B_1\over B_0}\pmb{Q}_0\hat{\pmb{C}}_1r{d\over dr}i\pmb{b}^T
=-{p\over 2p_{\rm mag}}Vc_1\bar\omega^2\hat{\pmb{C}}_0i\pmb{t}
-\left({B_1\over B_0}\pmb{Q}_0-{B_2\over B_0}\pmb{C}_0\right)\hat{\pmb{C}}_1i\pmb{b}^T,
\ee
and those for the solid crust are given by
\be
{B_1\over B_0}\pmb{Q}_1\hat{\pmb{C}}_0r{d\over dr}i\pmb{t}=
\hat{\pmb{C}}_1i\pmb{b}^T-\left[{B_1\over B_0}\left(2+{d\ln B_r\over d\ln r}\right)\pmb{Q}_1
-{B_2\over B_0}\pmb{C}_1\right]\hat{\pmb{C}}_0i\pmb{t},
\ee
\be
r{d\over dr}\pmb{W}=-\left(3-V\right)\pmb{W}
-\left[Vc_1\bar\omega^2\hat{\pmb{C}}_0+{\mu\over p}\left(2\hat{\pmb{C}}_0
+2\pmb{Q}_0\hat{\pmb{\Lambda}}_1
-\pmb{\Lambda}_0\hat{\pmb{C}}_0\right)\right]i\pmb{t}
+{2p_{\rm mag}\over p}\left[{B_1\over B_0}\left(2+{d\ln B_r\over d\ln r}\right)\pmb{Q}_0
+{B_2\over B_0}\pmb{C}_0\right]\hat{\pmb{C}}_1i\pmb{b}^T,
\ee
and
\be
\pmb{W}={\mu\over p}\hat{\pmb{C}}_0r{d\over dr}i\pmb{t}
+{2p_{\rm mag}\over p}{B_1\over B_0}\pmb{Q}_0\hat{\pmb{C}}_1i\pmb{b}^T,
\ee
where $i^2=-1$, and 
\be
V={GM_r\rho\over pr}, \quad c_1={(r/R)^3\over M_r/M}, \quad M_r=4\pi\int_0^r\rho r^2dr,
\ee
and $G$ is the gravitational constant, $M$ and $R$ are the gravitational mass and the radius
of the star, respectively, and
\be
B_1=B_r/\cos\theta, \quad B_2=-B_\theta/\sin\theta,
\ee
and
$p_{\rm mag}=B_0^2/8\pi$.
The definition of the matrices
$\pmb{C}_0$, $\pmb{C}_1$, $\pmb{Q}_0$, $\pmb{Q}_1$, $\pmb{\Lambda}_0$, and $\pmb{\Lambda}_1$ 
is found, for example, in Lee (1993), and
the matrices $\hat{\pmb{C}}_0$, $\hat{\pmb{C}}_1$, and $\hat{\pmb{\Lambda}}_1$ are defined by
\be
\hat{\pmb{C}}_0=\pmb{C}_0, \quad \hat{\pmb{C}}_1=\left\{\pmb{C}_1\right\}, 
\quad \hat{\pmb{\Lambda}}_1=\pmb{\Lambda}_1
\ee
for even modes, and
\be
\hat{\pmb{C}}_0=\left\{\pmb{C}_0\right\}, \quad \hat{\pmb{C}}_1=\pmb{C}_1, 
\quad \hat{\pmb{\Lambda}}_1=\left\{\pmb{\Lambda}_1\right\}
\ee
for odd modes, where $\left\{\pmb{A}\right\}=\left(A_{i,j+1}\right)$
for a matrix $\pmb{A}=(A_{ij})$.

For the inner boundary condition we require that the functions $i\pmb{t}$
and $i\pmb{b}^T/r$ are regular at the stellar center.
The boundary condition at the stellar surface is given by $i\pmb{b}^T=0$.
The jump conditions at the solid-fluid interfaces is the continuity of the function $i\pmb{t}$ and of
the $r\phi$ component of the traction, the latter of whcih leads to
\be
\left[{\mu\over p}\hat{\pmb{C}}_0r{d\over dr}i\pmb{t}
+2{p_{\rm mag}\over p}{B_1\over B_0}\pmb{Q}_0\hat{\pmb{C}}_1i\pmb{b}^T\right]_\pm=0,
\ee
where $[f(r)]_\pm=\lim_{\epsilon\rightarrow 0}\left[f(r+\epsilon)-f(r-\epsilon)\right]$.

\subsection{Neutron Star Models}

We integrate the Tolman-Oppenheimer-Volkoff equation (see, e.g., Shapiro \& Teukolsky 1983) 
to obtain zero temperature ($T=0$) neutron star models, where no effects of the magnetic field
on the equilibrium structure are included.
The equation of state used for the inner crust and the fluid core is that given by
Douchin \& Haensel (2001). The equation of state in the outer crust is that given by
Baym, Pethick, \& Sutherland (1971), and for the fluid ocean we simply use the equation of state
for a mixture of a completely degenerate electron gas and a non-degenerate gas of Fe nuclei. 
The boundary between the fluid ocean and
the outer crust is set rather arbitrarily at the density $\rho=10^4$g cm$^{-3}$.

For the solid crust, we employ the average shear modulus $\mu_{\rm eff}$ (Strohmyer et al 1991), 
which in the limit of $\Gamma\equiv(Ze)^2/(ak_BT)\rightarrow\infty$ is given by
\be
\mu_{\rm eff}=0.1194\times {(Ze)^2n\over a},
\ee
where $n$ is the number density of the nuclei and $a$ denotes the separation between the nuclei
defined by
\be
{4\pi\over 3}a^3n=1,
\ee
and $k_B$ is the Boltzmann constant.

For modal analyses, we calculate neutron star models of mass $M=1.245M_\odot$ and $M=0.8967M_\odot$.
For the former model, the radius and the central pressure and density are
$R=1.18\times10^6{\rm cm}$, $p_c=10^{35}$ dyne/cm$^2$, and $\rho_c=8.75\times10^{14}$ g/cm$^3$, 
respectively. 
The ratio of the thickness of the crust, $\Delta r_{\rm crust}$, to the radius is
$\Delta r_{\rm crust}/R\simeq 0.093$, and the surface ocean above the crust is very thin.
For the latter model, the parameters are $R=1.19\times10^6{\rm cm}$, 
$p_c=5\times 10^{34}$ dyne/cm$^2$, $\rho_c=6.75\times10^{14}$ g/cm$^3$, 
and $\Delta r_{\rm crust}/R\simeq 0.139$,
respectively.

Figure 1 plots the frequency $\omega$ 
of the crustal toroidal modes of the two models for low values of harmonic degree $l$, where
$\omega_0=\sqrt{GM/R^3}$ and $\nu=\omega/2\pi$, and no effects of magnetic field are included.
For each value of $l\ge 2$,
the lowest frequency mode is the fundamental mode that has no radial nodes
of the eigenfunction.
There exists no fundamental mode for $l=1$.
The figures show that
the fundamental mode frequency rapidly increases with increasing $l$, 
but the frequencies of the overtones remain almost constant for varying $l$.
As first discussed by Hansen \& Cioffi (1980), the frequency of the fundamental toroidal 
mode in the solid crust is rather insensitive to the neutron star mass, and
that of the overtones is inversely proportional to the thickness $\Delta r_{\rm crust}$ of the crust, 
which is consistent with Figure 1.

\begin{figure}
\resizebox{1.\columnwidth}{!}{
\includegraphics{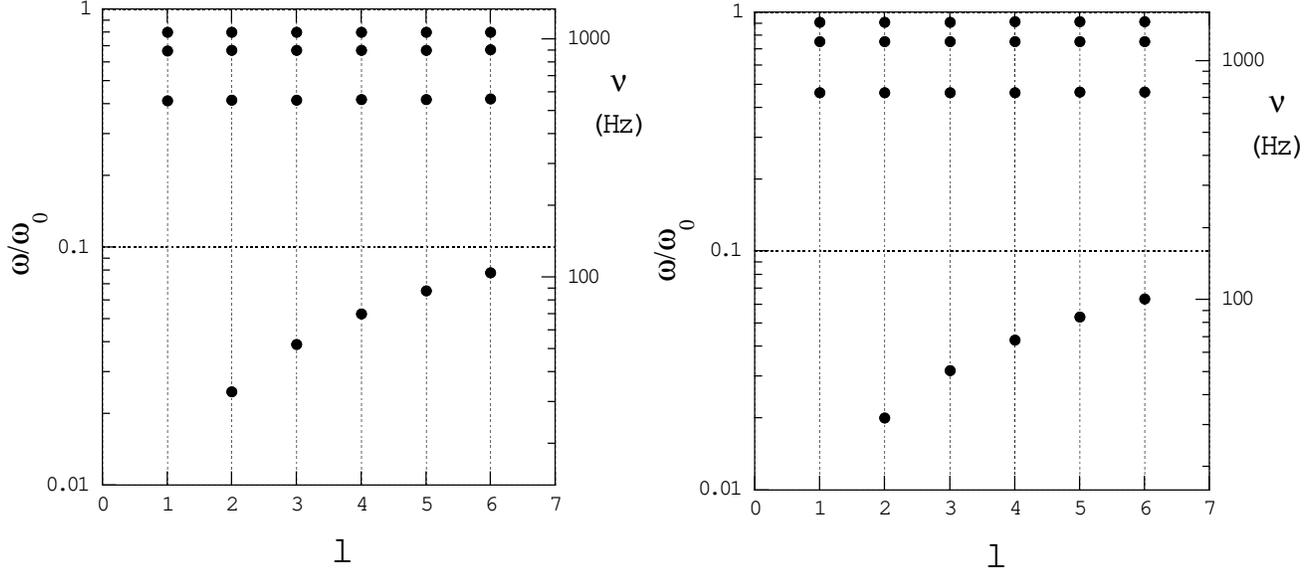}}
\caption{Frequencies of the crustal toroidal modes for low degree $l$s,
where $\omega_0=\sqrt{GM/R^3}$ and $\nu=\omega/2\pi$, and
no effects of magnetic fields are included to calculate the modes.
The left panel is for the $M=0.8967M_\odot$ model, and the right panel for the $M=1.245M_\odot$
model, respectively.
For each value of $l\ge2$, the lowest frequency mode is the fundamental mode having
no radial nodes of the eigenfunction. No fundamental mode exists for $l=1$.}
\end{figure}

\section{Numerical Results}

With our numerical method employed to calculate toroidal modes of a magnetized star, 
we find numerous solutions to the oscillation equations for a given $B_p$.
Most of the solutions thus obtained, however, are dependent on $j_{\max}$, the length of the expansions,
and we have picked up only the solutions that are independent of $j_{\max}$.

\begin{figure}
\resizebox{1.\columnwidth}{!}{
\includegraphics{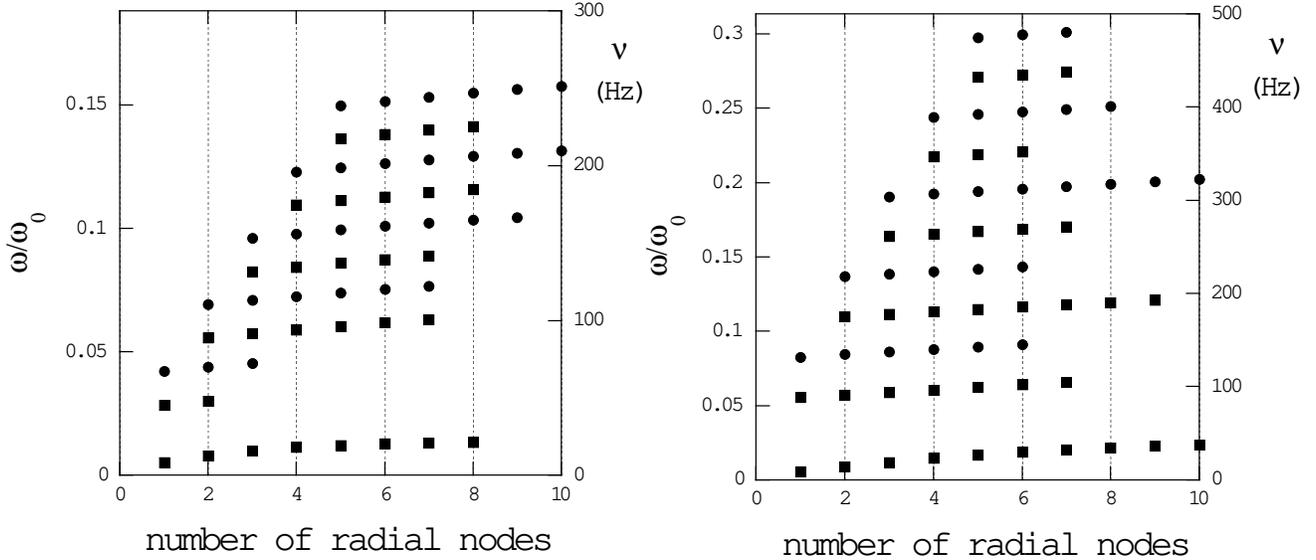}}
\caption{Frequencies of toroidal modes of the $M=1.245M_\odot$ model
versus the number of radial nodes of $iT_{l_1}$, where $\omega_0=\sqrt{GM/R^3}$ and $\nu=\omega/2\pi$,
and the filled squares and circles denote even and odd modes,
respectively.
The left panel is for the case of $B_p=5\times10^{15}$G, and the right panel for $B_p=10^{16}$G.
No fundamental modes without radial nodes are found.}
\end{figure}

\begin{figure}
\resizebox{0.5\columnwidth}{!}{
\includegraphics{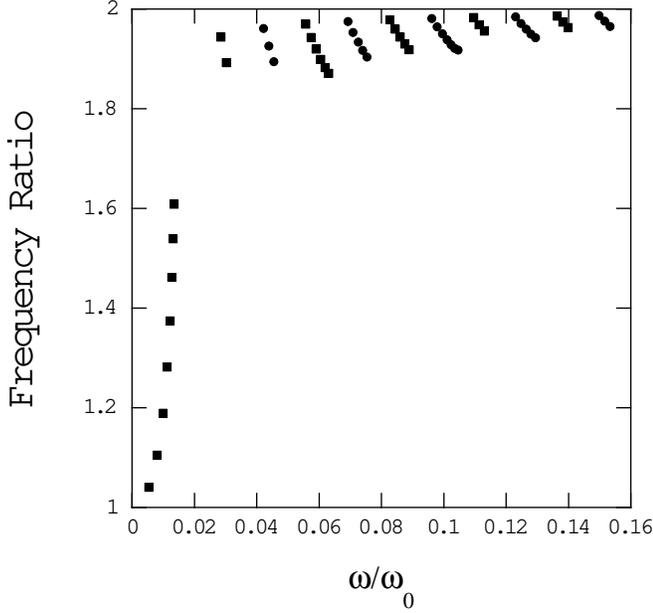}}
\caption{Frequency ratio $\omega(10^{16}{\rm G})/\omega(5\times10^{15}{\rm G})$ versus
$\omega(5\times10^{15}{\rm G})/\omega_0$ for the model of $M=1.245M_\odot$, 
where the filled squares and circles are
for even and odd modes, respectively.}
\end{figure}

\begin{figure}
\resizebox{1.\columnwidth}{!}{
\includegraphics{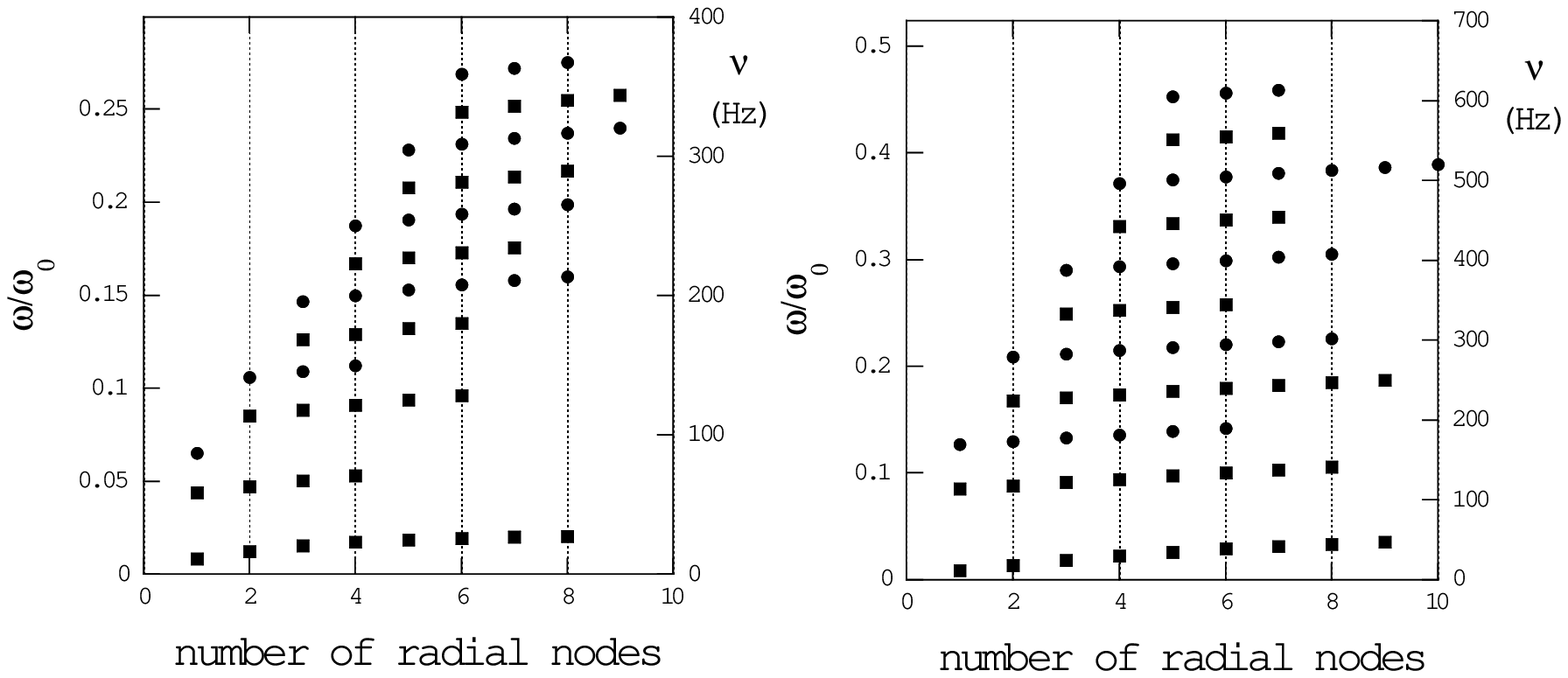}}
\caption{Same as Figure 2 but for $M=0.8967M_\odot$.}
\end{figure}

In Figure 2 the oscillation frequencies of the toroidal modes of the $M=1.245M_\odot$ model
are plotted versus the number of radial nodes of the expansion coefficient $iT_{l_1}$, 
where the left panel is for $B_p=5\times10^{15}$G and the right panel for $B_p=10^{16}$G, and
the filled circles and squares indicate odd and even modes, respectively.
The figures clearly show the existence of distinct mode sequences, in each of which
the frequency of the mode remains rather constant
and increases only slowly as the number of nodes increases.

To compare the two cases of $B_p=10^{16}$G and $B_p=5\times10^{15}$G, we plot in Figure 3
the frequency ratio $\omega(10^{16}{\rm G})/\omega(5\times10^{15}{\rm G})$ versus
$\omega(5\times10^{15}{\rm G})/\omega_0$, where the filled squares and circles are
for even and odd modes, respectively.
Except for the lowest frequency sequence of the even modes, the ratio is approximately
equal to $\sim 2$, and it decreases as the radial order increases, indicating that
the frequency is approximately proportional to the field strength and 
is largely determined by the strength of the magnetic field.
This is not the case for the lowest frequency sequence, for which the ratio rapidly
increases from $\sim 1$ to $\sim 2$ as the radial order increases, demonstrating that the response of
the oscillation frequency to the field strength is not the same as that for the overtone sequences.
We note that the lowest frequency mode found in the figure is a much more slowly increasing function
of $B_p$ than the modes in the overtone sequences.

It is convenient to write the mode frequency as
\be
\omega_p(k,n)= \omega_b+(k+j_p)\Delta\omega  + (n-k)\delta\omega+\delta(k,n),
\ee
where the non-negative integer $k$ denotes
the order of the mode sequences of a given parity (even or odd), and the integer $n(\ge k)$ 
is the number of nodes of the expansion coefficient $iT_l$, and the integer $j_p$ is
set equal to $0$ for the even mode sequences $(p=e)$ and to $1/2$ for the odd mode sequences $(p=o)$.
Since there exists no fundamental modes that have no radial nodes of the expansion
coefficients, we require $n\ge 1$ even for $k=0$, for which we have to replace $n$ with $n-1$ in equation (24).
The quantities $\omega_b$, which is set equal to the lowest frequency obtained,
$\Delta\omega$, and $\delta\omega$ are assumed independent of $k$ and $n$,
and the quantity $\delta(k,n)$ is assumed to be substantially small compared with $\Delta\omega$ and
$\delta\omega$.
Note also that we find no mode sequence having $k=0$ for odd modes, and hence we have to start with $k=1$
for them.
Except for the case that involves the lowest frequency sequence of even modes,
$\omega_p(k+1,n)-\omega_p(k,n)=\Delta\omega+\delta(k+1,n)-\delta(k,n)$ 
does not strongly depend on $k$ and $n$, and
$\omega_p(k,n+1)-\omega_p(k,n)=\delta\omega+\delta(k,n+1)-\delta(k,n)$ slowly decreases as $n$ increases.
It is interesting to note that the quantity $\Delta\omega$ is approximately proportional to
the field strength $B_p$, but the quantity $\omega_p(k,n+1)-\omega_p(k,n)$ is rather
insensitive to $B_p$.
For example, for the model of $M=1.245M_\odot$, we have $\Delta\omega/\omega_0\simeq 0.052$
for $B_p=10^{16}$G and $\Delta\omega/\omega_0\simeq 0.0255$
for $B_p=5\times10^{15}$G, and $\delta\omega/\omega_0\simeq0.0017$ for both cases.

Figure 4 is the same as Figure 2 but for the model of $M=0.8967M_\odot$, 
the crust of which is thicker than that of the $M=1.245M_\odot$ model.
Although the frequency spectra look almost the same between the two models,
for a given set of $(p,k,n)$ and $B_p$, the frequency found for the $M=0.8967M_\odot$ model is
higher than that for the $M=1.245M_\odot$ model.
This is because the fluid core radius $R-\Delta r_{\rm crust}$ (the Alfv\'en velocity $v_A\propto B_p/\sqrt{\rho}$)
of the $M=0.8967M_\odot$ model is smaller (larger) than that of the $M=1.245M_\odot$ model, and hence the 
traveling time $(R-\Delta r_{\rm crust})/v_A$ of magnetic perturbations in the core
for the former is shorter than for the latter.

As examples of the eigenfunctions of the toroidal modes, we plot the expansion coefficients 
$iT_{l_j^\prime}$ and $ib^T_{l_j}$ of three even modes as functions of the fractional radius
$x=r/R$ in Figures 5 to 7, 
where the solid, long-dahsed, short-dashed, and dotted lines are the expansion coefficients
with $j=1$ to 4, respectively.
For the lowest frequency sequence even modes plotted in Figures 5 and 6, 
the amplitudes $xiT_{l^\prime}$ are much larger than those of
$ib^T_l$, indicating that the kinetic energy is dominating the magnetic energy.
The magnetic perturbations $ib^T_l$ have amplitudes also in the solid crust.
This is not the case for the modes in the overtone mode sequences having $k\ge1$ as exemplified by Figure 7, 
where the amplitudes of the magnetic perturbations are much larger than $xiT_{l^\prime}$ and are
strongly confined in the fluid core, having almost no amplitudes in the crust.
Also note that the expansion coefficients with different values of $j$ seem to coalesce to
a single curve in the outer core of the star.

\begin{figure}
\resizebox{1.\columnwidth}{!}{
\includegraphics{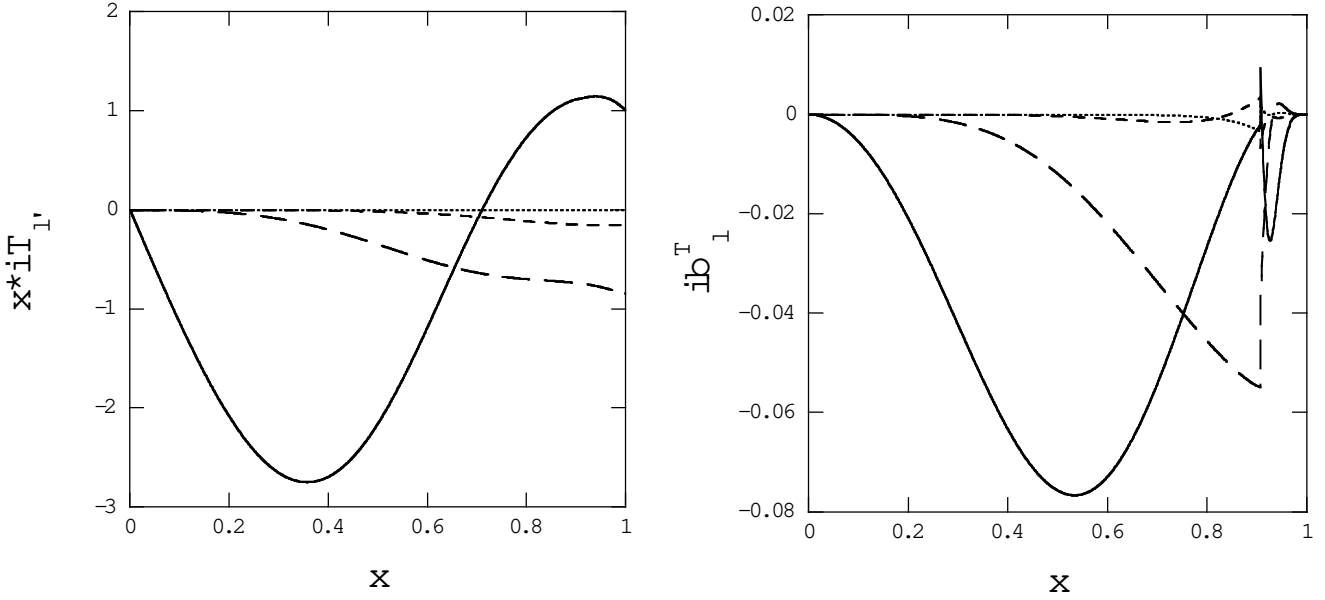}}
\caption{Expansion coefficients $x iT_{l^\prime_j}$ and $ib^T_{l_j}$ as functions of the fractional
radius $x=r/R$ for the even toroidal modes of $\omega_e(0,1)/\omega_0=0.005515$ 
for $B_p=10^{16}$G, where $\omega_0=\sqrt{GM/R^3}$ and $M=1.245M_\odot$.
The solid, long-dashed, short-dashed, and dotted lines are for the expansion coefficients
associated with $j=1$ to 4, respectively.
The amplitude normalization is given by $iT_{l^\prime_1}=1$ at the stellar surface $x=1$.}
\end{figure}
\begin{figure}
\resizebox{1.\columnwidth}{!}{
\includegraphics{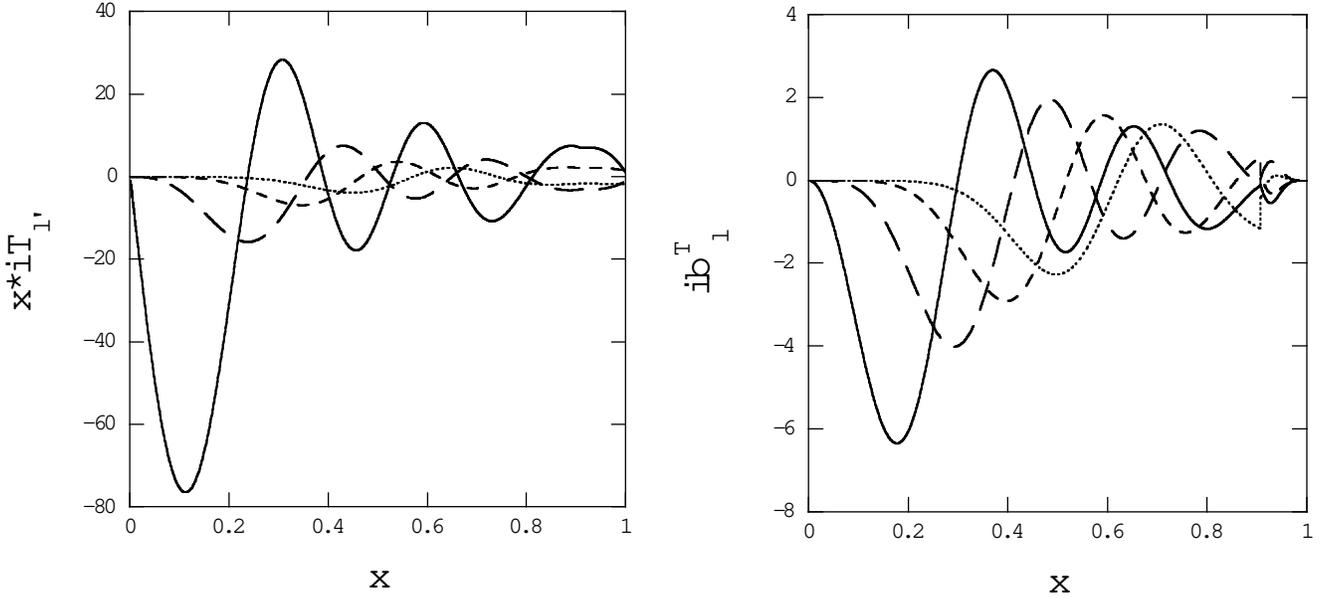}}
\caption{Same as Figure 5, but for the even mode of $\omega_e(0,5)/\omega_0=0.01671$.}
\end{figure}
\begin{figure}
\resizebox{1.\columnwidth}{!}{
\includegraphics{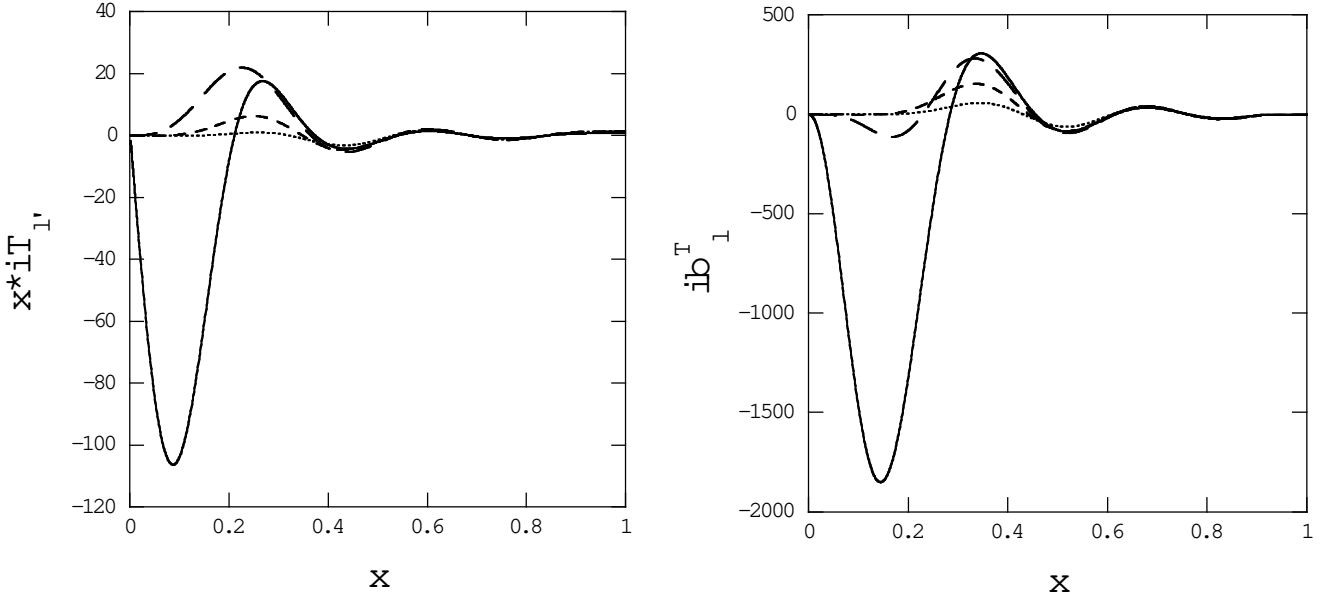}}
\caption{Same as Figure 5, but for the even mode of $\omega_e(5,5)/\omega_0=0.2709$.}
\end{figure}

Figures 8 to 10 give plots of the expansion coefficients $iT_{l^\prime}$ and $ib^T_l$
for three odd toroidal modes associated with
$(k,n)=(1,1)$, (1,5), and (5,5).
The magnetic perturbations $ib^T_l$ of the modes have much larger amplitudes
than $xiT_{l^\prime}$ and they are strongly confined in the fluid core.
The functions $xiT_{l^\prime}$ have comparable amplitudes both in the fluid core and
in the solid crust.
As in the case of even modes, the expansion coefficients of the mode (5,5)
tend to coalesce to a single curve in the outer core region.

\begin{figure}
\resizebox{1.\columnwidth}{!}{
\includegraphics{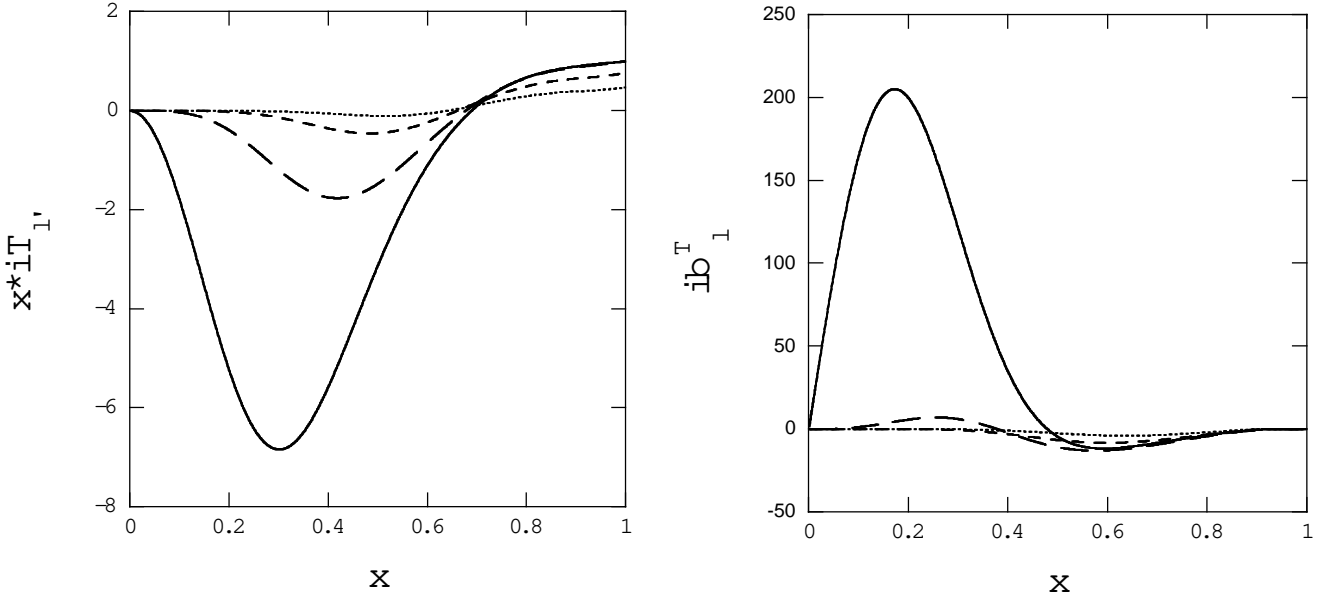}}
\caption{Expansion coefficients $x iT_{l^\prime_j}$ and $ib^T_{l_j}$ as functions of the fractional
radius $x=r/R$ for the odd toroidal modes of $\omega_o(1,1)/\omega_0=0.08271$
for $B_p=10^{16}$G, where $\omega_0=\sqrt{GM/R^3}$ and $M=1.245M_\odot$.
The solid, long-dashed, short-dashed, and dotted lines are for the expansion coefficients
associated with $j=1$ to 4, respectively.
The amplitude normalization is given by $iT_{l^\prime_1}=1$ at the stellar surface $x=1$.}
\end{figure}
\begin{figure}
\resizebox{1.\columnwidth}{!}{
\includegraphics{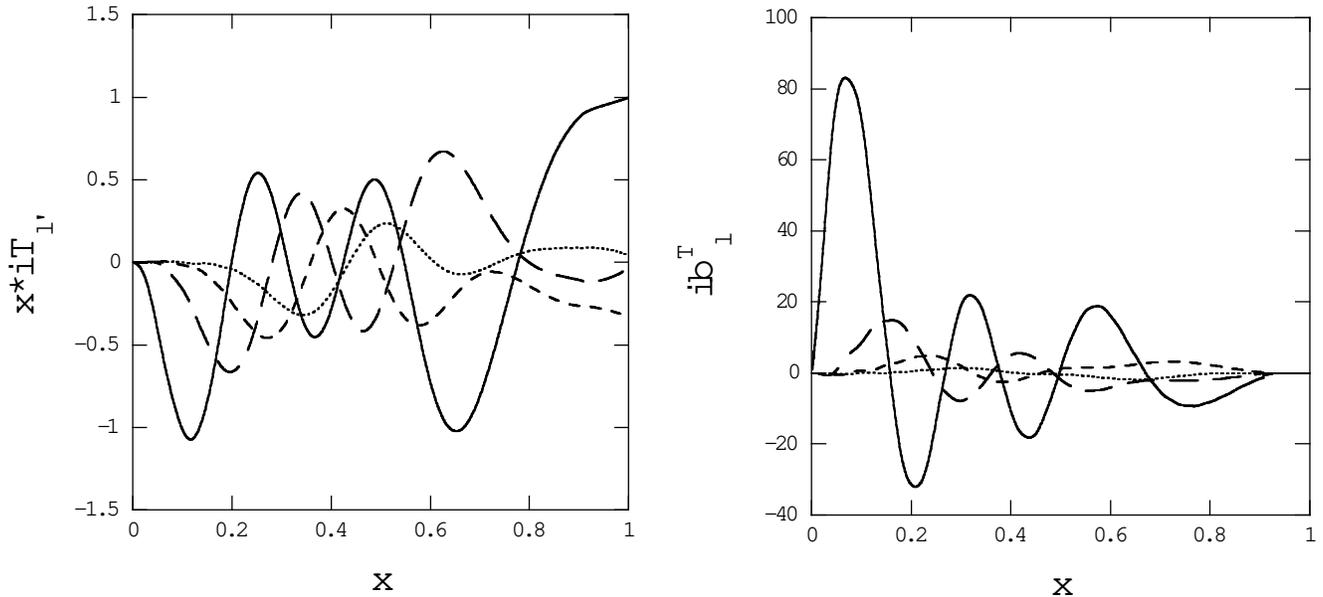}}
\caption{Same as Figure 8, but for the odd mode of $\omega_o(1,5)/\omega_0=0.08936$.}
\end{figure}
\begin{figure}
\resizebox{1.\columnwidth}{!}{
\includegraphics{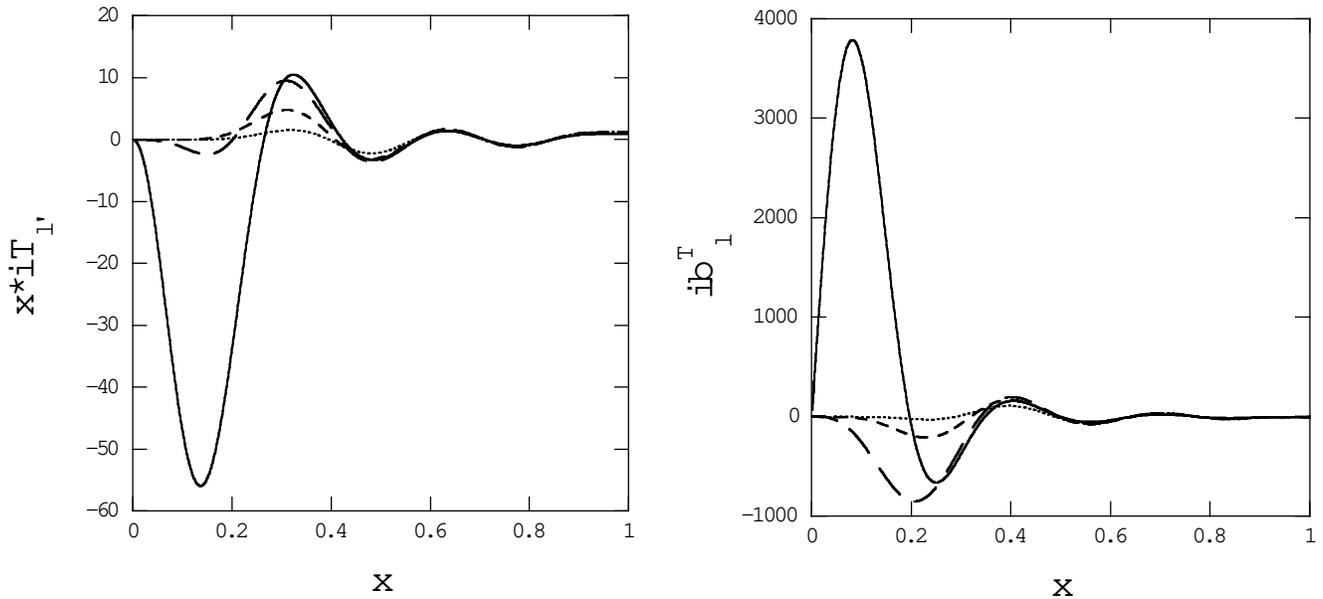}}
\caption{Same as Figure 8, but for the odd mode of $\omega_o(5,5)/\omega_0=0.2976$.}
\end{figure}

We need to examine how the existence of a solid crust affects
the frequency spectrum of the toroidal modes.
Figure 11 shows the toroidal mode frequencies of the $M=1.245M_\odot$ model
for $B_p=5\times10^{15}$G, where the modes are
calculated by treating the entire interior as a fluid.
Without the crust, the lowest frequency sequence is composed of odd modes, and
the even mode sequence corresponding to $k=0$,
which appears as the lowest sequence when the solid crust is included, is missing.
Therefore, for the fluid star we have to set $j_p$ equal to $j_p=0$ for odd mode sequences with $p=o$ and
to $j_p=1/2$ for even mode sequences with $p=e$ in equation (24).
In the lowest frequency odd mode sequence, there exists the fundamental mode
having no radial nodes of the expansion coefficients,
corresponding to $(k,n)=(0,0)$.
Since the oscillation equations (12) and (13) for fluid regions contain
the oscillation frequency only in the form of the ratio $\bar\omega/B_p$,
the frequency of a given $(p,k,n)$ is
exactly proportional to the field strength $B_p$.
This suggests that the quantities $\omega_b$,
$\Delta\omega$, $\delta\omega$, and $\delta(k,n)$ defined in equation (24)
are also proportional to $B_p$.
Note that if we write $\omega_b\propto B_p^\beta$ for the lowest frequency mode
for the case with a solid crust, 
where the exponent $\beta>0$ may be weakly dependent on $B_p$, 
we have $\beta\ltsim 1/4$ for $B_p\gtsim10^{15}$G, 
the feature of which is different from the case without the crust.
Figure 11 indicates that
the frequency separation between the lowest and second lowest sequences is 
approximately $\sim\Delta\omega/2$ for the case
without the crust, but this separation is given by $\sim\Delta\omega$ for the case with the crust.
Note that the frequency separation between the second and third lowest sequences, for example, 
is given by $\sim\Delta\omega/2$ for both cases.

\begin{figure}
\resizebox{0.5\columnwidth}{!}{
\includegraphics{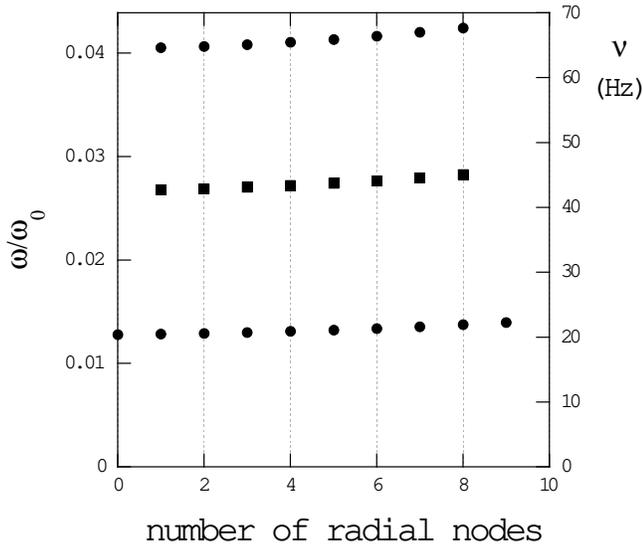}}
\caption{Frequencies of the toroidal modes of the $M=1.245M_\odot$ model versus the number of radial nodes of
the expansion coefficient $iT_{l^\prime_1}$ for $B_p=5\times10^{15}$G,
where $\omega_0=\sqrt{GM/R^3}$ and $\nu=\omega/2\pi$, and the filled squares and circles are for
even modes and odd modes, respectively.
The modes are calculated by treating the entire interior as a fluid.
}
\end{figure}

As suggested by Figures 2, 4, and 11, it is not always possible to numerically
find $j_{\rm max}$ independent modes associated with large values of $n$ and $k$.
Possible numerical reasons for the difficulty may be that the modes we want are immersed in a dense
spectrum of $j_{\max}$ dependent solutions, particularly for weak $B_p$, and that 
a large $j_{\max}$ and high numerical precision are required to
correctly calculate the expansion coefficients in the vicinity of the stellar center
for modes associated with large $n\ge k$.
From the theoretical point of view, however, it is not necessarily clear that the 
toroidal modes of the kind we find in this paper
exist for arbitrary sets of integers $k(\ge 0)$ and $n(\ge k)$.

\section{conclusions}

In this paper, we have calculated toroidal modes of magnetized stars with a solid crust, where
the entire interior of the star is assumed to be threaded by a poloidal magnetic field
that is continuous at the stellar surface to the outside dipole field.
We find distinct mode sequences of the toroidal modes, in each of which the mode frequency
remains rather constant and only slowly increases as the radial order of the modes increases.
In the presence of a solid crust, the frequency separation between the lowest and
second lowest frequency mode sequences
is approximately given by $\Delta\omega$, but that between the second and third lowest 
frequency mode sequences by $\Delta\omega/2$, where the frequency separation $\Delta\omega$ is
roughly proportional to the field strength $B_p$.
This frequency pattern of the low frequency sequences is different from that found
for the model without the crust, for which the frequency separation between the sequential sequences of
low frequency modes is given by $\Delta\omega/2$.
We also find that
for the equation of state we use, $\Delta\omega$ is larger for smaller $M$.

The eigenfunction $\xi_\phi$ of the modes belonging to the lowest frequency sequence have 
much larger amplitudes than $B^\prime_\phi$ and can penetrate into the solid crust.
On the other hand, $\xi_\phi$ of the modes belonging to the higher frequency sequences
is much smaller than $B^\prime_\phi$, which is well confined into the fluid core and
does not have any substantial amplitudes in the solid crust.

The frequency ranges of the toroidal modes we find for the magnetized neutron star
with $B_p\sim10^{16}$G overlap the QPO frequencies found for the magnetar candidates, SGR 1806-20
and SGR 1900+14.
This suggests that
we may interpret  the observed QPOs based on the magnetic toroidal modes, 
and that detailed comparisons between observed frequency spectra and theoretical calculations
make it relevant to infer physical parameters of the magnetar candidates, 
such as the equation of state and the strength of the magnetic field.
But, we think it worth pointing out that except for the modes belonging to the lowest frequency
sequence, the magnetic perturbations, which have much larger amplitude than $\xi_\phi$, 
are well confined in the fluid core and do not have substantial amplitudes in the solid crust,
which make it difficult for the modes to be directly observable.
If the magnetar candidates do no have a crust, the problem of observability
could be avoided.
In this case, however,
we have to use more realistic surface boundary conditions than $\pmb{B}^\prime=0$ used in the present
calculations.
Note that, although the existence of a solid crust affects the frequency pattern
of the low frequency mode sequences, the frequency range of the magnetic modes
itself is not very much dependent on the presence or absence of a solid crust unless the crust is 
extremely thick.

We have tried to find toroidal modes well confined in the solid crust or
core toroidal modes that are in resonance with the crustal toroidal modes, but failed.
We also find it extremely difficult to identify distinct toroidal mode sequences 
for magnetic fields weaker than $B_p\ltsim10^{15}$G when a solid crust is included in the models.
It is therefore not clear whether the toroidal mode sequences of the kind 
we find in the present paper
can survive also for weakly magnetized neutron stars with a solid crust.
If the field strength is much weaker than $B_p\sim10^{15}$G, the magnetic fields may have
only minor effects on the crustal toroidal modes (e.g., Lee 2007), and 
it will be justified to use frequency
spectra of the crust modes theoretically obtained in the weak field limit
to interpret observed QPOs.
As briefly noted in the last section,
it is not theoretically clear how the frequency spectra of the toroidal modes of magnetized stars 
should look like, and how
the toroidal modes behave in the limit of $n\ge k\rightarrow\infty$.
We may even speculate that the frequency spectra we obtained reflect the existence of
continuous spectra (see, e.g., Goedbloed \& Poedts 2004, see also Levin 2007), but the 
detailed numerical analysis of continuous frequency spectra is beyond the scope of this paper.
It will be worthwhile
to examine the effects of an interior toroidal field on magnetic modes.
It is also needed to extend the present analysis to a general relativistic formulation 
(e.g., Sotani et al 2006, 2007).

\end{document}